# Does economics need a scientific revolution?

Ivan O. Kitov, IDG RAS

**Abstract**

Economics does not need a scientific revolution. Economics needs accurate measurements according to high standards of natural sciences and meticulous work on revealing empirical relationships between measured variables.

The proclamation by J.-P.Bouchaud [1] of the necessity "*to break away from classical economics and develop completely different tools*" might please many researchers and definitely bought a strong ideological support among professionals in the hard sciences. Sharing the general mood of the proclamation it is worth to analyze it without any favor to the physical sciences and without prejudice to economics. There are several questions arising directly from the Jean-Philippe's text and some real problems behind the text reflecting the inability of the broad scientific community to build a sound economic theory matching strict requirements developed in empirical and quantitative disciplines.

First question is related to a striking tone of the pamphlet. For a physicist, it is very uncommon to use a somewhat proclamatory language and slightly negligible attitude to opponents. The strength of physics has been always expressed in the overall fit between predictions and observations. This fit does not need trumpets – just scatter plots. There is a flavor of total superiority in the wording about physics as the only source of all those useful things.

So, why this powerful science did not explain economic and financial crises yet?

Second question is a consequence of the failure to create a physics-like theory of economic evolution, as Bouchaud clearly evidences himself. Why should one make a blank shot? This attack from the side of natural sciences is easily suppressed by the first question and the dry residual is expressed by the cited economist: "*These concepts are so strong that they supersede any empirical observations*". This statement is 90 percent right because the opposite statement is not proved by available economic data. And this statement will be 100% wrong only when some new economic concept fits economic data, as physics has been demonstrating since Sir Isaac Newton. Meanwhile, the blank shot undermines respect to physics as a science and makes any opposition to the current status of the mainstream economics void. The best weapon against alchemy is analytic chemistry, but not words about analytic chemistry.

Third question is related to the actions required from the target audience of the essay. It seems that the rhetoric better complies with specialist from the hard sciences. What should those numerous researchers accomplish? The author gives a dangerous answer – "*completely different tools*". Therefore, he admits that there exists *no* tool capable to resolve most urgent problems of theoretical and experimental economics. This is another evidence of the inconsistency of physics in the realm of economics. On the other hand, this answer implies the abolishment of the most powerful part of the hard sciences – measurements.

A minor theoretical but a big practical agenda is formulated in the end of the essay. The author proposes to change the mindset of the specialists in economics and finances by the introduction more natural sciences in curricula. It is difficult to disagree that the diversity in the mandatory subjects to read allows a wider view on short- and long-term problems in any discipline. However, dozens thousands of young physicists, mostly theoreticians, in quant funds failed to prevent the current crisis. Hence, the strength of solid knowledge and the diversity of physical education is not a remedy. One should first give an appropriate physical concept of economy and then include it in education.

The above polemics would be void if no convincing examples of physics-like behavior of economic variables are presented. In a sense, there is a bunch of words against another bunch, with economists condescendingly smiling. However, before demonstrating couple robust micro- and macroeconomic relationships, it is obligatory to formulate and discuss the fundamental problem of economics as a science, which is echoed in all three questions above. So, the real problem is *data*.

It is a miracle that nor economists neither physicists (!) have focused on economic data as the only source and proof of economic concepts. The former have an excuse of the absence of any experience in handling model-producing data. For centuries, data were an alien in the field of economics, as openly expressed by all (I mean literally all) statistical agencies responsible for economic measurements. When publishing data, they always warn users that the data (real GDP, inflation, labor force, unemployment, productivity, etc.) are not compatible over time due to revisions to definitions and procedures. As a consequence, no economist had any chance to work with quality data of appropriate length. No experience – no regret. Prescott, the Nobel Prize winner in economics, wrote [2]:

> *… This raises the question of why inductive or empirical inference proved sterile in business cycle research. This sterility was not due to the incompetence of the researchers who pursued the inductive approach. The group who pursued this research program included a disproportionate number of the best minds in economics. The reason these inductive attempts failed, I think, is that the policy invariant laws governing the evolution of economic system is inconsistent with dynamic economic theory …*

It is more difficult to imagine the mental shift in numerous physicists who build theories based on wrong data. The habit to work with data of sufficient quality is so natural that they have never dug deep enough to recognize this basic mistake. There is no opportunity to build a reliable economic theory when data measured in different units are used together. The failure to develop empirically validated models of economic processes is an inevitable consequence of the absence of measurements matching standard requirements.

One must check data consistency before modelling, as we have done for several macroeconomic and demographic series. When checked and corrected, where possible, these data provide an invaluable source of information revealing numerous links between macroeconomic variables. Here only three examples are presented, from many. First is the model describing the evolution of personal income distribution (PID) over time and age, which is borrowed from geomechanics [3]. Left panel of Figure 1 depicts observed and predicted average income as a function of working experience in the United States for 1967 and 2001, the former is measured by the Census Bureau in the Annual Social and Economic Supplement of the Current Population Survey. Right panel of Figure 1 shows the evolution of Gini coefficient during the same period as measured and predicted by the model. This is the microeconomic level of description because the model accurately predicts the number of people with any given income or any given year after 1947 using only from real GDP per capita and age pyramid. It also predicts the observed evolution of income inequality. The model provides an adequate quantitative description of personal incomes and meets general requirements for physical models. It supports the idea that in economic terms society in the United States is a physical system.

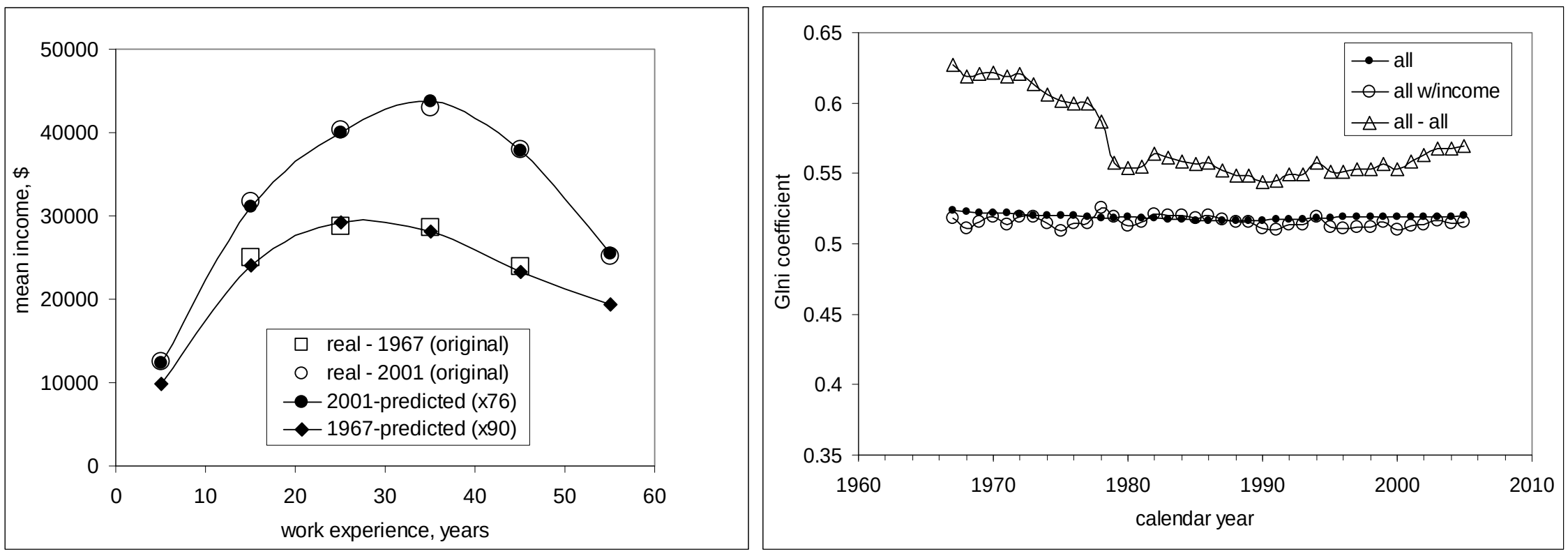

Figure 1. *Left Panel:* Comparison of observed and predicted mean personal income dependence on work experience in 1967 and 2001. Averaging in 10-year intervals of work experience. *Right panel*: Comparison of predicted (solid squares) and empirical (open triangle - all working age population; open circle – only people with income) Gini coefficient between 1967 and 2005.

Second example is from the field of macroeconomics and represents a part of a comprehensive macro-model for the US economy [4], which explains the evolution of macroeconomic variables only by the change in population, i.e. in the number of people in a given country. Under this framework, the pair inflation/unemployment is driven by the change in the level of labor force. Figure 2 demonstrates the robustness of this link for inflation in the United States, with inflation lagged by 10 quarters behind the change in labor force. Annual readings of inflation in the left panel of Figure 2 are shifted back by 2.5 years in order to synchronize them with the change in labor force. Smoothing of the latter curve provides a much better fit between the curves at the level of $R^2$>0.9 and still allows prediction a several quarters horizon. Workforce projections developed by the CBO or BLS allow forecasting at a horizon of several years – the U.S. should expect a deflationary period from 2012. This deflation is caused by low rate of labor force growth and thus is similar to that observed in Japan. A helpful consequence of the link is the possibility to replace the measurements of inflation by the measurement in labor force. Right panel of Figure 2 displays cumulative curves for those in the left panel. The difference between the predicted and observed cumulative curves is a stationary or an I(0) process [5], i.e. the difference cumulates to zero over time. Considering fundamentally different nature of these two variables, one might use relationship:

$$\pi(t) = 4.5dLF(t-2.5)/LF(t-2.5) - 0.031$$

where $\pi(t)$ is the CPI inflation at time $t$, $LF(t-2.5)$ is the level of labor force 10 quarters before, as an empirical law in economics. It is similar to the estimation of the distribution of density in the Earth from orbits of satellites.

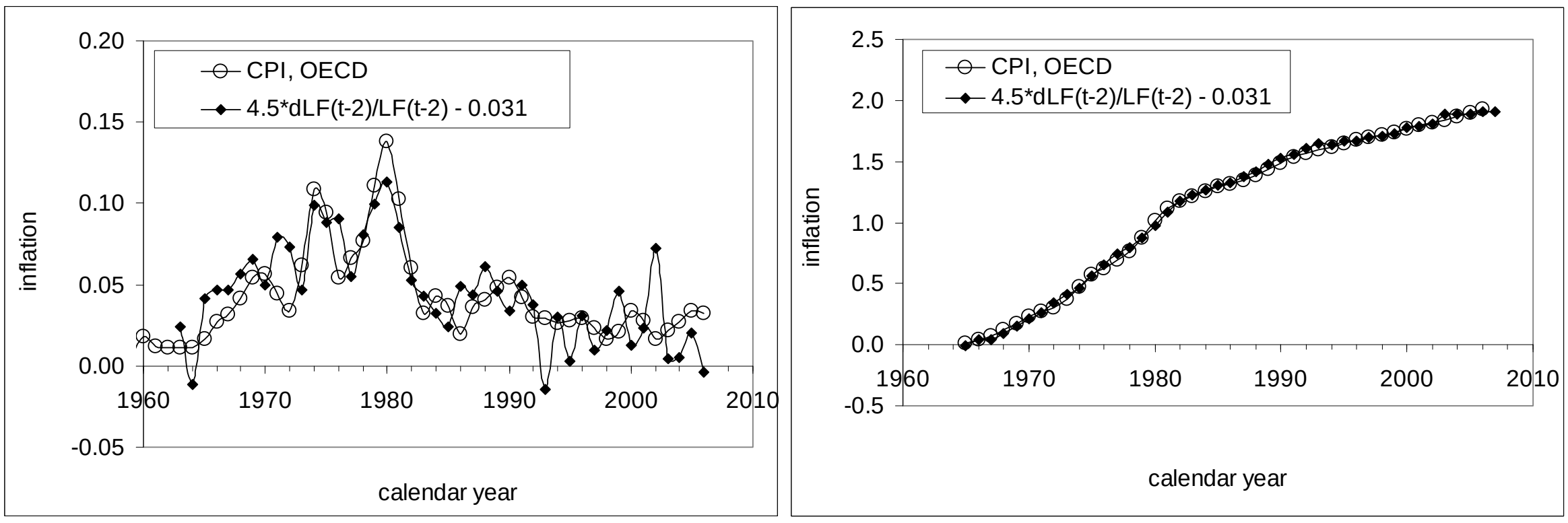

Figure 2. Observed and predicted (CPI) inflation in the United States. *Left panel* – annual rate curves. *Right panel*: cumulative curves.

Finally, unemployment in Italy provides an extraordinary example of the time delay between the change in labor force and unemployment [5]. The lag is the largest determined so far, with 6 years in the USA. Figure 3 depicts observed unemployment and that predicted from labor force according the following relationship:

$$UE(t) = 3.0dLF(t-11)/LF(t-11) + 0.085$$

where $UE(t)$ is the unemployment at time $t$, $LF(t-11)$ is the level of labor force 11 years before. Because of large fluctuations in the change rate of labor force this time series is smoothed with a 5-year moving average, MA(5), which still allow an out-of-sample forecast at a 9-year horizon with RMS forecasting error (RMSFE) of 0.55%. This is an outstanding result, which also a decisive example for the validation of the model. Here we get our feet back on the ground of empirically driven sciences: data reveal a reliable ($R^2$=0.92) relationship between two measured variables. The relationship predicts at a very long horizon with an uncertainty much lower than the expected change. The horizon is lengthy but the reward is high – robust empirical bounds in economics.

All three examples, and dozens more not shown here for the sake of brevity, are completely data-driven. There was no theoretical assumption or a desire to develop a revolutionary tool behind the obtained relations. The model for personal incomes is the result of a random attempt to apply a concept specific for geomechanics to some measured value in economics, with as long time series as possible. The intuition behind the macro-model for developed economies is trivial – dirty trial-and-error with the simplest linear link, potentially lagged.

Therefore, there is no demand for revolution in economics. One desperately needs a significant increase in the quality of economic data, the data being measured according to sound definitions of macro-economic variables under study. As in physics, statistical inferences are only possible then the length and volume of quality data will reach the threshold, which is well established in the hard sciences. Hence, the way economics should walk along is to repeat the loop data-model-data, which has been productive for centuries in the natural sciences. On this path, actual revolution is not feasible, because there is nothing revolutionary in the repetition, with small deviations, of what other people have already done. One needs the meticulous everyday work and progressive correction of the attained knowledge, just as in physics. Then, real revolution is economic behavior of society and individuals is likely to come.

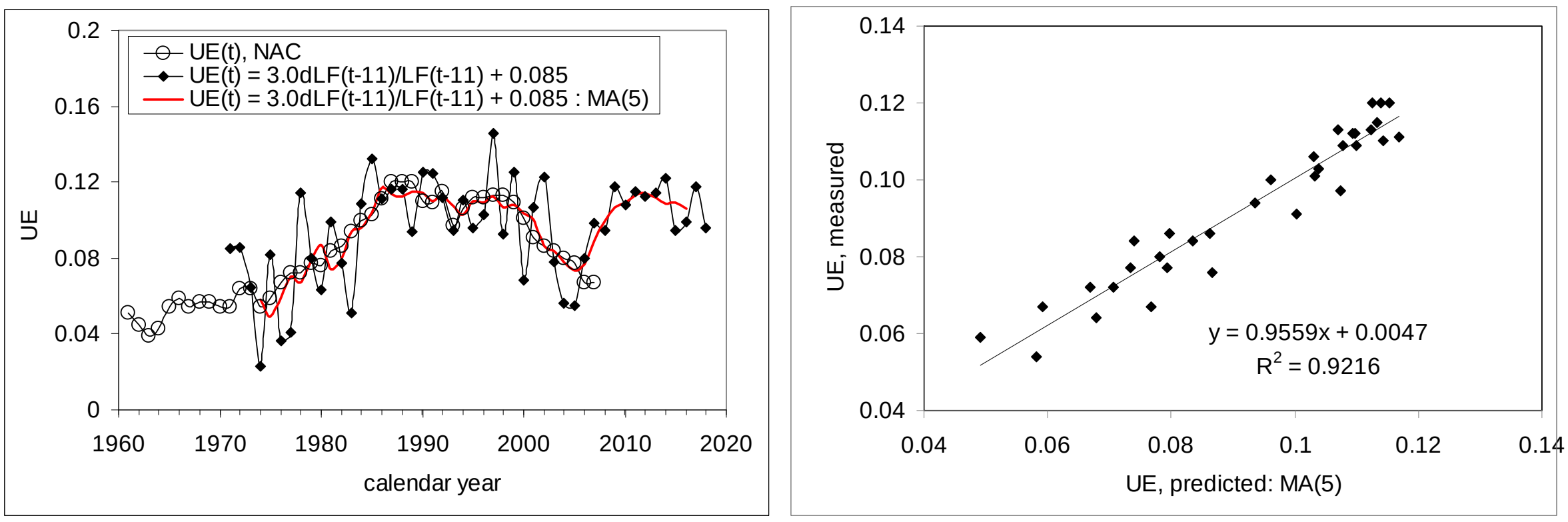

Figure 3. Observed and predicted unemployment in Italy. The prediction horizon is 11 years. Due to large fluctuations associated with measurement errors in the labor force the predicted series is smoothed with MA(5). Linear regression gives $R^2$=0.92 for the period between 1973 and 2006, with RMSFE of 0.55%. The unemployment should start to increase in 2008.